# Generation of arbitrary complex quasi-non-diffracting optical patterns

Antonio Ortiz-Ambriz,[1] Servando Lopez-Aguayo,[1] Yaroslav V. Kartashov,[2,3] Victor A. Vysloukh,[4] Dmitri Petrov,[2] Hipolito Garcia-Gracia,[1] Julio C. Gutiérrez-Vega[1], and Lluis Torner[2]

[1]*Photonics and Mathematical Optics Group, Tecnólogico de Monterrey, Monterrey, México 64849*
[2]*ICFO-Institut de Ciencies Fotoniques, and Universitat Politecnica de Catalunya, Mediterranean Technology Park, 08860 Castelldefels (Barcelona), Spain*
[3]*Institute of Spectroscopy, Russian Academy of Sciences, Troitsk, Moscow region, 142190, Russia*
[4]*Departamento de Fisica y Matematicas, Universidad de las Americas – Puebla, 72820, Puebla, Mexico*

**Abstract:** Due to their unique ability to maintain an intensity distribution upon propagation, non-diffracting light fields are used extensively in various areas of science, including optical tweezers, nonlinear optics and quantum optics, in applications where complex transverse field distributions are required. However, the number and type of rigorously non-diffracting beams is severely limited because their symmetry is dictated by one of the coordinate system where the Helmholtz equation governing beam propagation is separable. Here, we demonstrate a powerful technique that allows the generation of a rich variety of quasi-non-diffracting optical beams featuring nearly arbitrary intensity distributions in the transverse plane. These can be readily engineered via modifications of the angular spectrum of the beam in order to meet the requirements of particular applications. Such beams are not rigorously non-diffracting but they maintain their shape over large distances, which may be tuned by varying the width of the angular spectrum. We report the generation of unique spiral patterns and patterns involving arbitrary combinations of truncated harmonic, Bessel, Mathieu, or parabolic beams occupying different spatial domains. Optical trapping experiments illustrate the opto-mechanical properties of such beams.

## 1. Introduction

Propagation-invariant light fields maintaining their transverse intensity distribution over indefinitely long distances and possessing unique self-healing properties play an extremely important role in various areas of science and are employed in diverse applications on both micro- and nano-scales [1]. In nonlinear optics, non-diffracting beams are utilized for efficient second-harmonic beam generation [2] and for beam shaping and steering in optically induced lattices in photorefractive nonlinear crystals [3-8]. Also, they provide confinement and strongly affect the evolution of matter waves in Bose-Einstein condensates [9,10], and several applications of light-induced lattices in quantum computing [11,12] and quantum optics in general [13,14] are being exploited. Non-diffracting beams have been also employed successfully for trapping and sorting micro-particles [15,16], their controllable transfer [17-19] between different micro-fluidic chambers and for the transfer of angular momentum from light beams to matter [20]. Such beams can be used for the creation of optical lattices for multi-particle and colloidal studies [21,22] and in biophysics applications, for example, for guiding neuronal growth [23-25] and studying molecular motors [26,27]. In all of these applications the specific transverse intensity distribution of the non-diffracting light field and the distance over which the beam can be considered invariable are the key factors determining the behavior of the system. Thus, in experiments on optical tweezers, the transverse intensity distribution determines the gradient forces acting on the micro-particles and their final arrangement [15]. In nonlinear optics and Bose-Einstein condensates the optical lattices induced by non-diffracting beams determine the spectrum of linear propagating eigenmodes and the symmetry of stationary nonlinear states [7,8]. This has inspired a quest for light patterns that remain invariable in the course of propagation over the distance required in each class of experiments and which pose spatial shapes with new symmetries; thereby, affording new opportunities.

The existence of a specific class of linear solutions of the scalar Helmholtz equation representing non-diffracting light fields has attracted considerable attention since the publication of the seminal work of Durnin, Miceli and Eberly on Bessel beams [28,29]. The wave-vectors of all plane waves (spatial harmonics) constituting any rigorously non-diffracting beam are located on the cone of fixed angle and all these vectors have identical magnitudes [1]. Consequently, plane waves forming the beam do not experience dephasing upon propagation and the beam does not suffer any distortions and might even self-heal in the course of propagation if the input ideal field distribution was locally perturbed. However, most non-diffracting light patterns available to date correspond to the known sets of exact solutions of the scalar Helmholtz equation. Using group theory, Kalnins and Miller [30] demonstrated that there are only four coordinate systems in which the scalar Helmholtz equation is separable, yielding invariant solutions along the propagation axis: plane waves in Cartesian coordinates, Bessel beams in circular cylindrical coordinates

[28,29], Mathieu beams in elliptic cylindrical coordinates [31,32] and parabolic beams in parabolic cylindrical coordinates [33,34]. In addition, there are asymmetric Airy beams [35,36] that remain diffraction-free but accelerate in the transverse plane in the course of propagation. The beams associated with each of the coordinate systems mentioned above feature specific unique symmetry (thus, the intensity distribution of Bessel beams contain multiple concentric rings, while parabolic beams are curved in the transverse plane).

However, in various areas, including optical tweezers, nonlinear optics and the physics of matter waves, the generation of patterns propagating without distortions over considerable distances and featuring symmetries beyond circular, elliptic, or parabolic ones may be required. Recently, we predicted theoretically that nearly arbitrary quasi-non-diffracting beams that do not belong to the above mentioned four fundamental beam families, could be obtained by engineering an angular spectrum of the desired kernel field distribution [37,38]. In this approach, a slight broadening of the spectrum of the beam, beyond the ideal infinitely narrow ring, allows a drastic increase of the complexity of spatial beam shape at the expense of the introduction of a slow diffraction broadening, whose rate is proportional to the angular spectral width. It should be mentioned that a similar approach based on sets of Helmholtz-Gauss beams, whose spectrum is characterized by a support of annular shape in the frequency space, was used in [39] for the experimental demonstration of complex patterns that remained quasi-invariant over the distance dictated by the width of the Gaussian envelope superimposed on the entire beam. However, the connection between the complexity of the spatial shape, the width of the angular spectrum of the beam, and the propagation length oven which the beam remains undistorted were not discussed. In this work we report the experimental verification of the method of generation of quasi-non-diffracting beams suggested by us in [37] and demonstrate various patterns whose symmetries drastically differ from symmetries of conventional rigorously non-diffracting Bessel, parabolic, or Mathieu beams. We experimentally and theoretically analyze the propagation of all generated patterns and illustrate that there is a direct relation between the complexity of the spatial shape of the beam, the propagation distance over which beams remain undistorted and the width of their ring-like spatial spectrum. We show the power of the technique by generating a number of quasi-non-diffracting patterns never obtained before, including spiraling beams and beams featuring completely different symmetries in different spatial domains. We illustrate the potential of such beams by using them for trapping micro-particles.

## 2. Construction of quasi-non-diffracting beams

The field of any *truly* non-diffracting beam propagating parallel to the $\xi$-axis of a uniform transparent material can be expressed via the Whittaker integral [30-34]:

$$q(\eta,\zeta,\xi) = \exp(-ik_\xi \xi) \int_0^{2\pi} G(\varphi) \exp[ik_t(\eta \cos\varphi + \zeta \sin\varphi)] d\varphi. \tag{1}$$

Here, $k_\xi$ and $k_t$ are the longitudinal and transverse components of the total wave number $k = (k_\xi^2 + k_t^2)^{1/2}$, respectively, $\varphi$ is the azimuthal angle in the frequency space, $\eta, \zeta$ and $\xi$ are the dimensionless transverse and longitudinal coordinates, respectively and $G(\varphi)$ is the angular spectrum that is defined along a circular delta-function of radius $k_t$ in the frequency space. By direct substitution, Eq. (1) can be verified as the diffraction-free solution of the Helmholtz equation

$$\frac{\partial^2 q}{\partial \xi^2} + \frac{\partial^2 q}{\partial \eta^2} + \frac{\partial^2 q}{\partial \zeta^2} + k^2 q = 0, \tag{2}$$

for *any* functional shape of infinitely narrow angular spectrum $G(\varphi)$. The characteristic transverse scale of such a beam is dictated by the parameter $k_t$ - a smaller $k_t$ corresponds to beams with larger features. By considering different angular spectra $G(\varphi)$, one may obtain non-diffracting fields that have explicit analytical expression, as in the case of the families of the four fundamental beams mentioned above, or fields that can be obtained only by means of numerical integration, as in the case of fractional [40] and random [41] non-diffracting beams.

While the construction of a non-diffracting beam using a known angular spectrum is straightforward, the inverse problem of the derivation of an angular spectrum for the pattern with a desired complex shape is much more complicated and as a rule, it is impossible to obtain a spectrum localized on a circular delta-function in frequency space. In [37,38] we found that even a slight broadening of the angular spectrum by $\delta k_t$ allows the generation of almost arbitrarily complex beams that remain nearly invariable upon propagation. In order to construct quasi-non-diffracting beams we use an iterative Fourier method, reminiscent of the methods employed in phase retrieval and image processing algorithms [42-44] and select the target field modulus $|q_{tg}(\eta,\zeta)|$ at $\xi=0$, whereas for the phase profile we use a quasi-random initial distribution. We perform a Fourier transform of this initial pattern and set to zero all components of the spectrum that fall outside the annular ring of width $\delta k_t$ and radius $k_t$. After this step, an inverse Fourier transform is calculated and the new field modulus is replaced with the desirable distribution $|q_{tg}(\eta,\zeta)|$ while retaining the new phase distribution. This procedure is repeated until convergence is achieved for a selected $\delta k_t$. The field from the last iteration is used without replacing its modulus with $|q_{tg}(\eta,\zeta)|$, such that some distortions always appear.

Importantly, the broadening of the angular spectrum by $\delta k_t$ results in a slow diffraction of the beam because upon propagation, the spatial harmonics in the beam experience slightly different phase shifts. For example, if a broad envelope with the width $w$ is superimposed on a non-diffracting beam, its angular spectrum acquires immediately the finite width $\delta k_t \sim 4w^{-1}$ but the beam can still be considered as non-diffracting over the distance $\sim wk/k_t$ [45,46]. Similarly, the rate of diffraction of quasi-non-diffracting beams constructed with an iterative algorithm remains very low as long as $\delta k_t \ll k_t$. The finite width of the angular spectrum introduces the second large characteristic transverse scale $1/\delta k_t$ of the beam, in addition to $1/k_t$ that sets the scale of the small details in the beam.

## 3. Experimental generation of quasi-non-diffracting patterns

Multiple approaches to the generation of propagation-invariant beams with various transverse intensity distributions are known (see [47,48]), including an approach based on masked Gaussian beams with annular support [49]. Many of them rely on spatial light modulators (SLMs). In our experiments, in order to convert the broad beam of the laser into a quasi-non-diffracting beam, we encoded the amplitude and phase distributions corresponding to the desired quasi-non-diffracting beam produced by the iterative Fourier algorithm, in a phase-only computer-generated hologram, by means of a spatial light modulator with a $800 \times 600$ resolution. The SLM is backlit by a collimated plane wave with $I \approx 20 \text{ mW/cm}^2$ from a low-power He-Ne laser operating at $\lambda = 632.8$ nm (see Fig. 1). The positive lens with $f_2 = 50$ cm placed at the distance $\xi = f_2$ from the SLM, allows us to observe the angular spectrum of the modulated pattern in its second focal plane. All the light not associated with the first diffraction order is filtered out by using a diaphragm. The diaphragm is placed in the focal plane of the second lens with $f_3 = 20$ cm, which provides an inverse Fourier transform of the spectrum and generates the desired quasi-non-diffracting beam at the distance $\xi = f_3$ behind the second lens. The intensity distributions in quasi-non-diffracting beams were detected using a CCD camera at various propagation distances with a step of $10$ cm. Our experimental setup described above is different from the approach used in [39], where it was necessary to use a checkerboard pattern to avoid zero diffraction order contribution at the Fourier plane. In our case a titled plane wave in a plane perpendicular to the propagation axis was added to the desired quasi-non-diffracting pattern upon construction of the computer-generated hologram. As a result, the first and zero diffraction orders propagate at different angles and zero order can be filtered out by using a diaphragm.

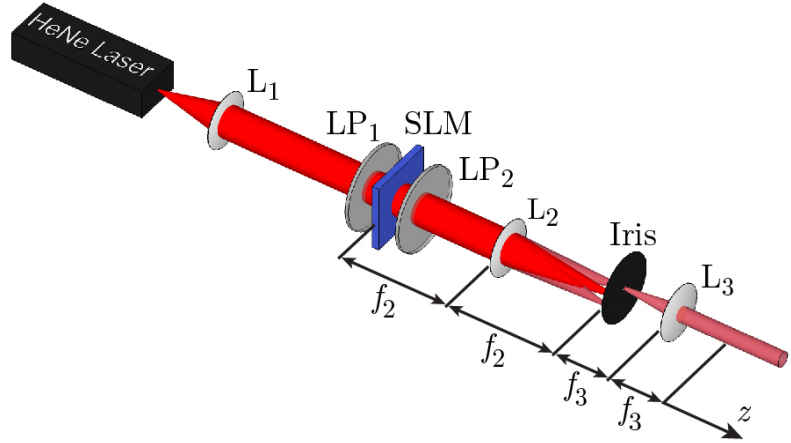

Fig. 1 Experimental setup used to generate quasi-arbitrary non-diffracting optical beams using a Spatial Light Modulator (SLM). The setup includes a collimating lens ($L_1$), two linear polarizers ($LP_1$ and $LP_2$) surrounding the SLM to block any residual phase modulation in the SLM, the input lens of the filtering stage ($L_2$ with focal distance $f_2$), the iris diaphragm used to filter the zero-order diffraction on the Fourier plane of $L_2$, and the output lens ($L_3$, focal length $f_3$). The propagation distance of the quasi-non-diffracting beam is measured from a focal distance $f_3$ away from the output lens $L_3$.

## 4. Angular spectrum of quasi-non-diffracting patterns and their diffraction rate

The impact of the width of the angular spectrum of the beam on its shape and distance, at which structural stability of the beam is conserved, is illustrated in Fig. 2 for the example of a specific spiraling beam. The first line illustrates the theoretically obtained spiraling beams for different spectrum widths $\delta = \delta k_t / k_t$, while the subsequent lines show their propagation in the experiment. One can observe that while the deviations of the input shape from the target ideal spiral intensity distribution are most pronounced for the narrowest angular spectrum $\delta = 0.1$ [Fig. 2(a)], the corresponding beam demonstrates an impressive and high structural stability upon propagation, nearly maintaining its unusual spiraling shape. The spiral pattern with $\delta = 0.2$ also maintains its shape upon evolution with only small deformations observable at 60 cm [Fig. 2(b)]. However, the beam with $\delta = 0.3$, which is quite close to the ideal spiral, exhibits fast rupture of the structure in the center and blurring of the spiral arms [Fig. 2(c)]. Notice that in all the beams in Fig. 2 the deformation develops mainly in the center, while the outer regions remain almost unaffected. This is in contrast to behavior of truncated non-diffracting beams, where the perturbation usually travels from the periphery to the beam center.

In order to show how the distance of diffraction-free propagation is related to the width of the ring-like angular spectrum, in Fig. 3 we show experimentally observed angular spectra corresponding to spiral patterns from Fig. 2. A wider spectrum corresponding to larger $\delta$ implies faster dephasing of spatial harmonics, i.e., faster diffraction. Recently, a similar connection between the distance over which the finite-energy Airy beams remain diffraction-free and their respective sharply truncated spectrum has been theoretically shown in [50]. Notice that for $\delta = 0.1$ and $0.2$ the amplitude of the angular spectrum is strongly modulated in the azimuthal direction, as occurs also for many truly non-diffracting patterns [Fig. 3(a) and 3(b)]. The spectrum of all spiral beams is modulated in the radial direction, which is especially obvious in Fig. 3(b), where one can observe considerable variations in the width of the ring. This observation suggests that the generation of patterns with clear spiraling shapes requires harmonics with different $k_t$ components and most likely, it will be impossible to find counterparts of these beams in the class of truly non-diffracting beams. Thus, our technique of angular spectrum engineering allows the observation of patterns that have no analogues among known non-diffracting beams.

To confirm that the beams shown in Fig. 2 are indeed quasi-non-diffracting beams, it is useful to compare their propagation with the propagation of localized beams whose width coincides with the characteristic scale $(\sim 1/k_\text{t})$ of the localized elements in the quasi-non-diffracting beam profiles, as was done for Bessel beams in [51]. Recall that in most practical applications where non-diffracting beams have been used, the most important observations were associated with their

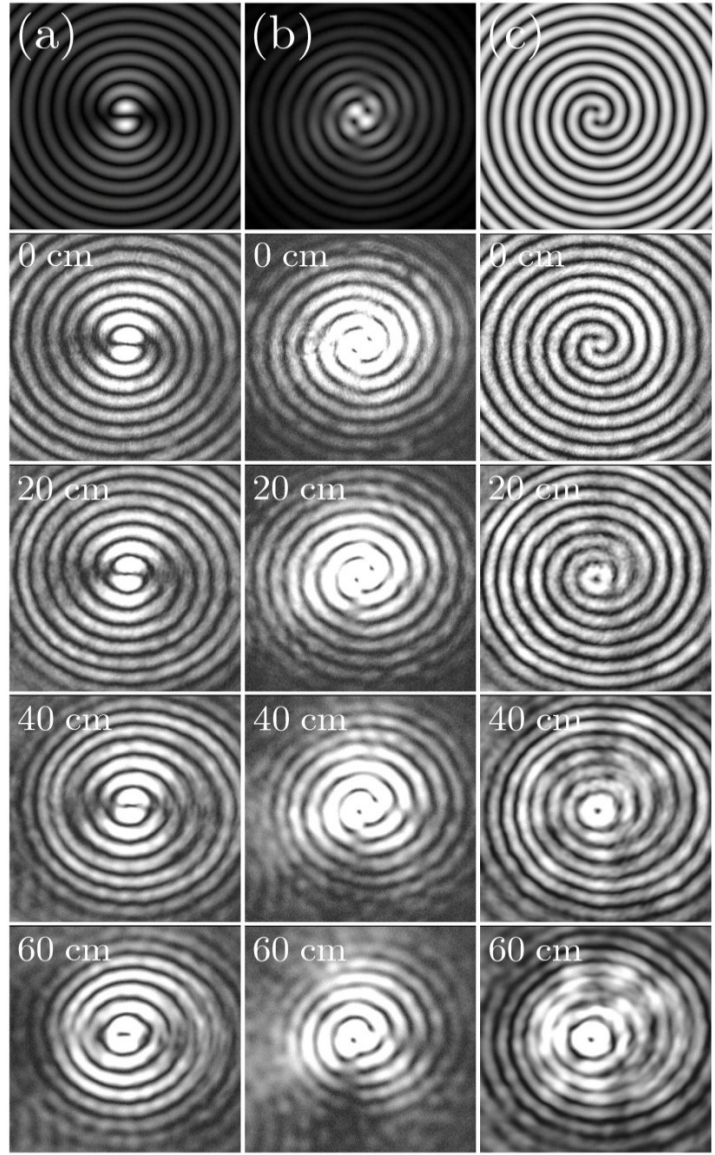

Fig. 2. First line shows quasi-non-diffracting spiraling beams generated numerically for different widths of angular spectrum $\delta = 0.1$ (a), $\delta = 0.2$ (b) and $\delta = 0.3$ (c). Second to fifth lines illustrate propagation of corresponding experimentally generated patterns.

small-scale structure, rather than with the broad envelope; thus, in tweezers or nonlinear optics, the particles or nonlinear beams are captured on domains with local intensity maxima. In the experiment with spiral patterns, the separation between neighboring spiral arms amounts to $r \approx 100~\mu\text{m}$, which gives the characteristic diffraction length $L_\text{dif} = \pi r^2 / \lambda$ of the order of

5 cm. Indeed, Gaussian or Laguerre-Gaussian beams with widths corresponding to $r \approx 100$ μm broaden dramatically after 60 cm propagation, corresponding to approximately 12 diffraction lengths (see experimental images in Fig. 4(a) and 4(b)), whereas spiral patterns with $\delta = 0.1$ were only slightly distorted at this propagation distance [Fig. 2(a)].

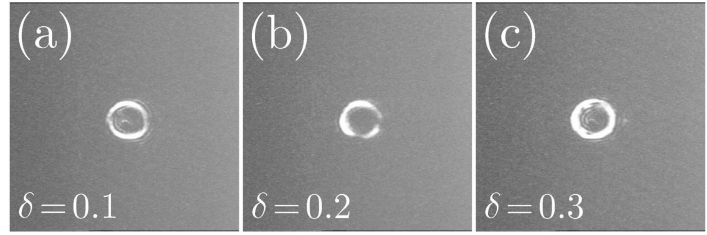

Fig. 3. Experimental angular spectra (obtained from first diffraction orders in focal plane of a lens) corresponding to spiral patterns in Fig. 2.

It should be stressed that while the angular spectrum width remains the main factor determining the distance of diffraction-free propagation of the desired pattern, its internal structure and symmetry are also crucial and sometimes different patterns corresponding to similar $\delta$ values might exhibit different degrees of structural distortion at the same propagation distance. Indeed, our iterative procedure only filters out all spatial frequencies located outside the annular ring of width $\delta k_t$, but under appropriate conditions the effective width of the spectrum of the beam obtained after iterations may be even *smaller* than the predetermined $\delta k_t$ value. Such beam would propagate undistorted for longer distance than the pattern whose spectrum extends over the entire allowed spatial frequency band $[k_t - \delta k_t/2; k_t + \delta k_t/2]$. This becomes especially obvious upon comparison of the evolution of the complex spiraling pattern from Fig. 2(a), whose spectrum occupies nearly the entire allowed range $\delta = 0.1$, with the evolution of truncated Bessel and Mathieu beams in Fig. 5, whose spectrum is effectively narrower for the same $\delta = 0.1$ value. The enhanced robustness of distributions from Fig. 5 is obvious. It should be also mentioned that the convergence of the iterative procedure strongly depends on the selected $\delta$ value. While for $\delta \to 0$ convergence is achieved only for a limited set of patterns that are very close to the exact non-diffracting beams with known symmetries, already for $\delta = 0.1$ the beams with almost any desired intensity distribution can be generated with a moderate ($<100$) number of iterations.

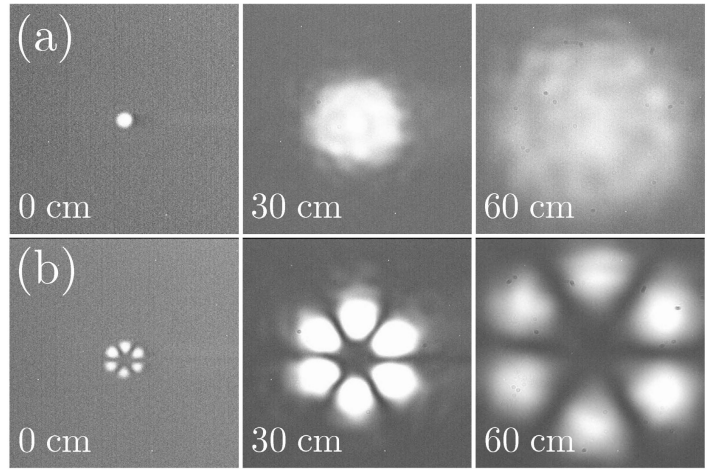

Fig. 4. Experimentally observed evolution of Gaussian (a) and Laguerre-Gaussian (b) beams with characteristic width $r_0 \sim 0.1$ mm.

## 5. Self-healing properties and complex quasi-non-diffracting optical patterns

In addition to diffraction-free propagation, another prominent property of non-diffracting beams is the ability to self-heal upon evolution after being perturbed. If such a beam is partially obstructed by an object, the transverse light pattern will tend to reconstruct itself after several diffraction lengths, if the perturbation is not too strong [52-54]. For this reason, it is not straightforward to nest defects into shapes of non-diffracting patterns that would be able to propagate unchanged. Such controllable deformations could be created easily with quasi-non-diffracting beams. The observation of quasi-non-diffracting patterns resembling Bessel and Mathieu beams with removed angular sectors is presented in Fig. 5(a) and 5(b), respectively. In contrast to exact truncated non-diffracting beams that would nearly self-heal at $60 \ cm$ of propagation, iteratively generated quasi-non-diffracting beams maintain induced defects. Moreover, these patterns survive for notably longer distances than the spiral beams shown in Fig. 1 and for them, the iterative procedure converges much faster, even for narrow angular spectra with $\delta \sim 0.05$. Similar results were obtained for other types of induced defects. Fig. 5(c) shows the propagation of another type of spiral beam that is also more robust than the objects shown in Fig. 2. Notice that the patterns presented in Fig. 5(b) and 5(c) are presented here for the first time.

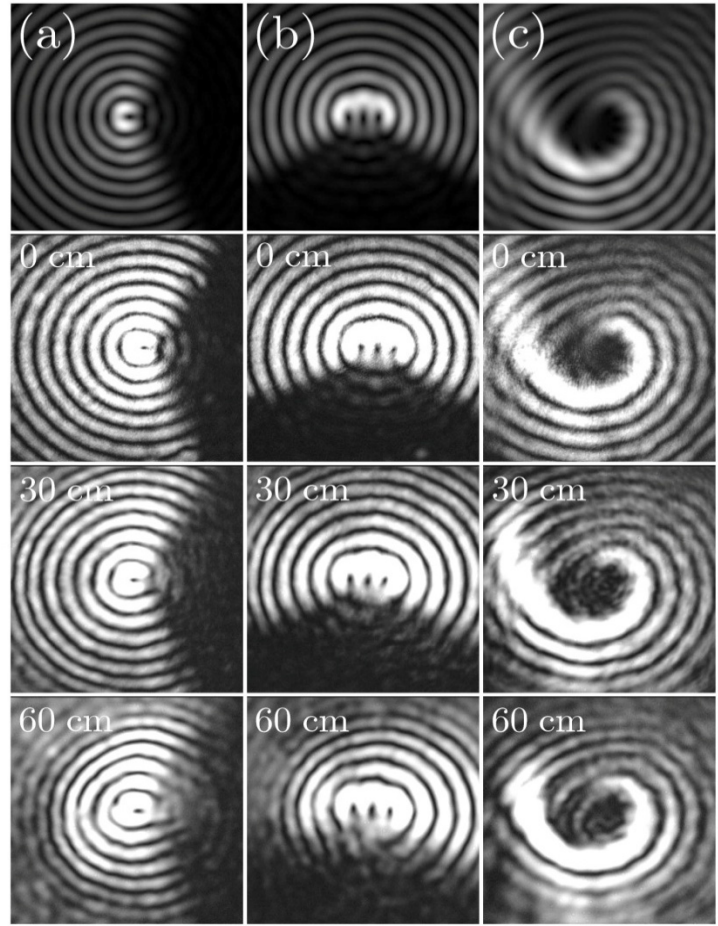

Fig. 5. First line shows quasi-non-diffracting Bessel (a) and Mathieu (b) beams with removed sectors, as well as spiraling pattern (c) generated numerically for angular spectrum width $\delta = 0.1$. Second to fourth lines illustrate propagation of corresponding experimentally generated patterns.

The method of angular spectrum engineering allows the generation of very specific patterns that feature completely different structures in different spatial domains and that can be observed experimentally. The evolution of such combined parabolic-Bessel and parabolic-cosine beams is shown in Fig. 6(a) and 6(b), respectively. Far from the transition region, where two beams are glued together and the intensity is strongly modulated, such patterns transform into corresponding representatives of truly non-diffracting beam families. These patterns also experience a remarkably slow deformation upon propagation; in Fig. 6, no pronounced deformations of such beams are visible even at 60 cm. It should be stressed that the experimental observations are completely consistent with the results of numerical simulations. As an example, we show the numerical simulations of the propagation of the parabolic-cosine beam in Fig. 6(c), which should be compared with the experimental distributions shown in Fig. 6(b). The first panel in this column shows the highly nontrivial phase distribution in the quasi-non-diffracting parabolic-cosine beam that is obtained during the iterative Fourier procedure.

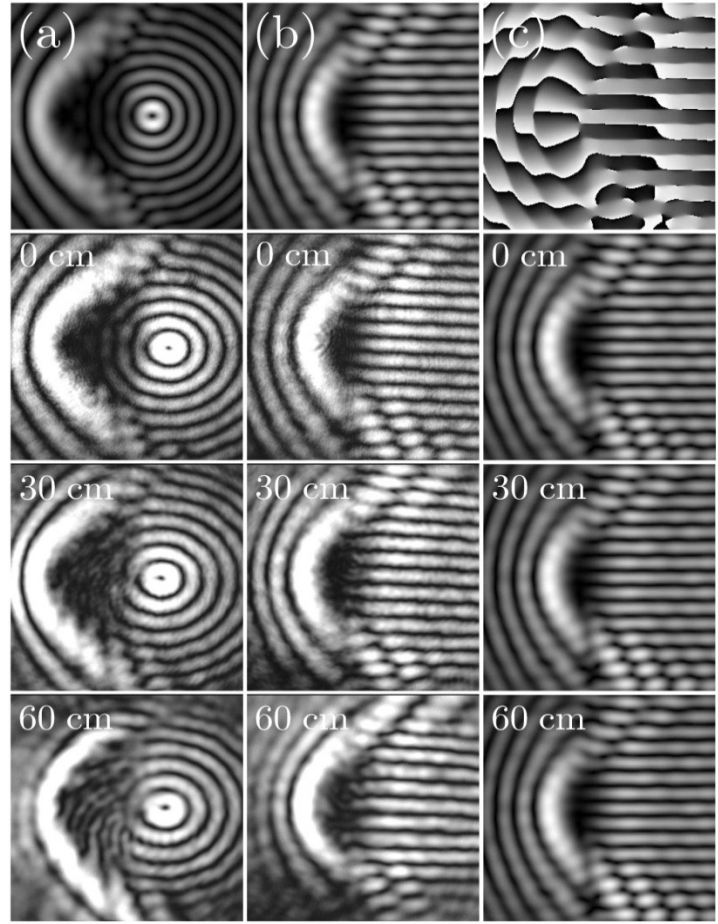

Fig. 6. (a) Parabolic-Bessel and (b) parabolic-cosine quasi-non-diffracting beams with different symmetries in right and left half-planes. First line shows numerically constructed beams for angular spectrum width $\delta = 0.1$. Second to fourth lines illustrate propagation of corresponding experimentally generated patterns. (c) Phase distribution in parabolic-cosine beam from (b) and numerical simulations of its propagation.

## 6. Trapping of micro-particles

We illustrate the properties of the quasi-non-diffracting beams by providing results on optical trapping. As non-diffracting beams by their nature have no variation of intensity in the axial ($\xi$) direction [15,16], no axial gradient can be created; thus, the optical trapping is restricted to two dimensions in the lateral plane. We first squeezed the parabolic-cosine beam by a lens system to a beam of about 30 $\mu$m in diameter. The beam then propagated through a fluid chamber filled with 2 and 3 $\mu$m polystyrene spheres. The fluid chamber was placed at an inverted microscope and therefore, the behavior of the spheres could be observed directly. The radiation pressure of the beam produces forces in the axial direction towards the cover slip of the fluid chamber, which immobilizes the spheres in the axial direction. Due to the Brownian motion, the spheres move randomly over the chamber in the absence of the beam but as Fig. 7 shows, the spheres of both diameters are concentrated near maxima in the intensity distribution in the presence of the quasi-non-diffracting beam and their positions only slightly fluctuate with time.

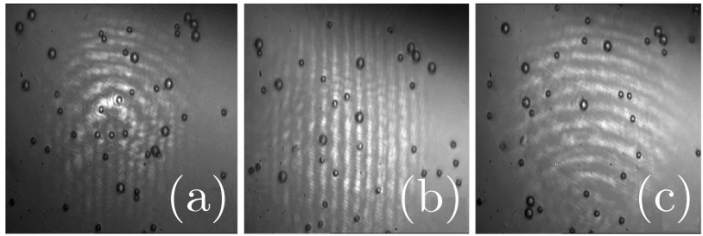

Fig. 7. Polystyrene spheres distributed in the focal plane of the imaging microscope objective when (a) the whole parabolic-cosine beam passes through the optical trapping system and when only the parts of the parabolic-cosine beam that contain straight lines (b) or curves (c) pass through the pinhole.

## 7. Conclusions

We reported the experimental demonstration of a powerful technique to generate a rich variety of quasi-non-diffracting optical patterns whose intensity distributions can be made nearly arbitrarily complex by engineering the angular spectrum of the beam. We showed that there exists a trade-off between the length of diffraction-free propagation, where the beam maintains its internal structure, and the deviation of the transverse beam shape from the ideal target pattern, e.g., spiral. The distance of diffraction-free propagation could exceed tens of diffraction lengths for sufficiently narrow angular spectra.

The beams presented here could be used in diverse areas of science, for example, for the formation of traps for individual neutral atoms [55] or for the control of matter waves hold in the corresponding optically-induced lattices. They could potentially be utilized in fluorescence microscopy where extended depth of the field is required [56]. Importantly, our approach may be readily extended to electron beams [57-59], opening up the corresponding important opportunities in electron microscopy and lithography.

## Acknowledgments

We acknowledge the financial support from the Consejo Nacional de Ciencia y Tecnología (Grant 182005), the Tecnológico de Monterrey (Grant CAT141) and also from Fundacio Privada Cellex Barcelona and the Generalitat de Catalunya (Grant 2009-SGR-159).